\documentclass[useAMS]{mn2e}

\usepackage{times}
\usepackage{psfig}

\def\spose#1{\hbox to 0pt{#1\hss}}
\newcommand\lsim{\mathrel{\spose{\lower 3pt\hbox{$\mathchar"218$}}
     \raise 2.0pt\hbox{$\mathchar"13C$}}}
\newcommand\gsim{\mathrel{\spose{\lower 3pt\hbox{$\mathchar"218$}}
     \raise 2.0pt\hbox{$\mathchar"13E$}}}

\title[3C66B as a TeV radiogalaxy] {3C66B as a TeV radio-galaxy}

\author[F. Tavecchio \& G. Ghisellini] 
{
Fabrizio Tavecchio\thanks{E--mail: 
fabrizio.tavecchio@brera.inaf.it} and Gabriele Ghisellini\\
INAF/Osservatorio Astronomico di Brera, via E. Bianchi 46, I--23807
Merate, Italy
}

\begin{document}


\pagerange{\pageref{firstpage}--\pageref{lastpage}} \pubyear{2008}

\maketitle

\label{firstpage}

\begin{abstract}
The MAGIC collaboration reported the detection of a new VHE source,
MAGIC J0223+430, located close to the position of the blazar 3C66A,
considered a candidate TeV blazar since a long time. A careful
analysis showed that the events with energies above 150 GeV are
centered on the position of the FRI radiogalaxy 3C66B (at 6 arcmin
from 3C66A), with a probability of 95.4\% (85.4\% including systematic
uncertainties) that the source is not related to 3C66A. We present a
model for the possible emission of 3C66B based on the structured jet
model already used to interpret the TeV emission of the radiogalaxy
M87. The model requires parameters similar to those used for M87 but a
larger luminosity for the layer, to account for the more luminous TeV
emission. We also show that the spectrum obtained by MAGIC can be
interpreted as the combined emission of 3C66B, dominating above
$\sim$200 GeV, and of 3C66A. The high--energy emission from the latter
source, being strongly attenuated by the interaction with the
extragalactic background light, can only contribute at low energies.
If we were to see the jet emission of 3C 66B at small viewing angles
we would see a spectral energy distribution closely resembling the one
of S5 0716+714, a typical blazar.
\end{abstract}

\section{Introduction}

Blazars represent the great majority of extragalactic objects detected
in the very high energy (VHE, $E>30$ GeV) $\gamma$--ray band
(Aharonian et al. 2008, De Angelis, Mansutti \& Persic 2008 for recent
reviews)\footnote{see also {\tt
http://www.mppmu.mpg.de/$\sim$rwagner/sources}}. Until now only one
non--blazar source, the nearby (16 Mpc) FRI radiogalaxy M87 is an
established VHE source (Aharonian et al. 2003, 2006a, Acciari et
al. 2008, Albert et al. 2008a). Due to the limited spatial resolution
of Cherenkov telescopes it is not possible to identify the TeV
emission region of M87.  However, the detection of variability on
short timescales ($\sim$ days, Aharonian et al. 2006a, Albert et
al. 2008a) suggests a compact emission region, possibly related to
knot HST-1 (e.g. Cheung, Harris \& Stawarz 2007), to the black hole
horizon (Neronov \& Aharonian 2007) or to the innermost regions of the
relativistic jet considered for blazars (Georganopoulos, Perlman \&
Kazanas 2005, Lenain et al. 2008, Tavecchio \& Ghisellini 2008,
hereafter TG08).

Recently, the MAGIC collaboration (Aliu et al. 2008) reported the
detection, during observations taking place between August and
December 2007, of a new source, MAGIC J0223+430, located close to the
position of the BL Lac object 3C66A. This blazar (with probable
redshift $z=0.444$, Miller et al. 1978, but see Finke et al. 2008) has
been considered a TeV candidate since a long time (Neshpor et
al. 1998, Stepanyan et al. 2002) and it has been recently detected by
VERITAS (with a soft spectrum, Swordy et al. 2008). However, a careful
analysis of the MAGIC data shows that the events with energies above
$\sim$150 GeV are rather centered on the position of the near (6
arcmin) FRI radiogalaxy 3C66B (located at 85.5 Mpc), with a
probability of 95.4\% (decreasing to 85.4\% if systematic
uncertainties on the positioning are taken into account) that MAGIC
J0223+430 is not related to 3C66A. The association with 3C66B is also
indirectly supported by the relatively hard measured spectrum (photon
index $\Gamma =3.1\pm 0.3$), difficult to reconcile with the expected
strong absorption of VHE photons, through interaction with the
extragalactic background light, for a source at the redshift af
3C66A. However, contamination from 3C66A (especially at low
$\gamma$--ray energies) cannot be completely excluded. If the
association of the excess with 3C66B will be confirmed, this would be
the second known TeV emitting radiogalaxy, indicating that M87 is not
an exceptional case.

In this letter we assume that the VHE emission detected by MAGIC comes
from 3C66B and propose (Section 2) that it is produced, as in the case
of M87, in the misaligned inner structured jet (Ghisellini, Tavecchio
\& Chiaberge 2005, hereafter GTC05, TG08).  For ``structured jet'' we mean a jet
composed by a fast internal spine surrounded by a slower layer. Both
the spine and the layer emit and there is a strong radiative interplay
between the two, since one component sees the radiation emitted by the
other relativistically boosted by the relative velocity.  In aligned
sources (i.e. blazars) we preferentially see the radiation produced by
the spine. Instead, for misaligned sources, as radiogalaxies, we tend
to see the radiation produced by the layer, since its lower bulk
Lorentz factor implies that its radiation is collimated within a
larger solid angle. As in the case of M87, we show that, if observed
under a small viewing angle, the SED of 3C66B would resemble that of a
typical BL Lac object, S5 0716+714, also recently detected in TeV band
by MAGIC (Teshima et al. 2008).  In Section 3 we discuss the possible
contribution of the (strongly attenuated) TeV emission of 3C66A to the
observed MAGIC spectrum.  Hereafter we assume the following
cosmological parameters: $H_{\rm 0}\rm =70\; km\; s^{-1}\; Mpc^{-1} $,
$\Omega_{\Lambda}=0.7$, $\Omega_{\rm M} = 0.3$.

\section{VHE emission of 3C66B from a misaligned structured jet}

\subsection{The model}

\begin{figure}
\vskip -1.3 cm
\centerline{ 
\hskip 1.4 cm
\psfig{file=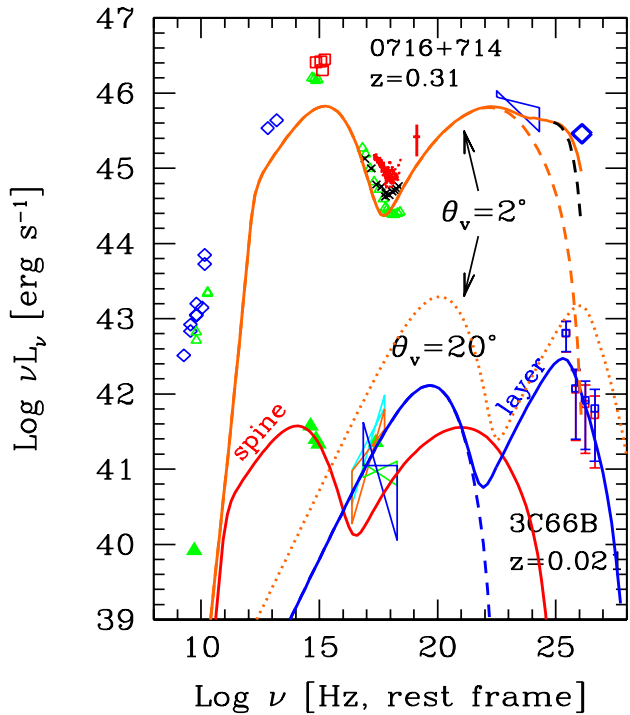,height=12.5cm,width=13.3cm} }
\vskip -1.3 cm
\caption{
SED of the core of 3C66B (lower part), 
represented with the equivalent isotropic luminosity as a 
function of the rest frame frequency. 
Radio through UV data (green filled triangles) are from
Trussoni et al. (2003) and references therein.  X--ray data (bow-ties)
are from Trussoni et al. (2003) (cyan and orange), Balmaverde et
al. (2006) (blue) and Evans et al. (2006) (green). MAGIC data (red
open squares) have been corrected for absorption (blue open squares)
with the LowSFR model of Kneieke et al. (2004). The curves correspond
to the structured jet model, considering the spine (red) and the layer
(blue) emission.  We also report (upper part) the emission seen by an
observer almost aligned with the jet axis ($\theta_{\rm v}=2^\circ$)
with (solid) and without (dashed) the contribution of the seed photons
of the layer for the scattering by electrons in the spine.
For comparison
we also report historical data (taken from NED, Tagliaferri et
al. 2003, Foschini et al. 2006, Hartman et al. 1999) describing the
SED of the BL Lac S5 0716+714. The blue open diamond indicates the TeV
flux as detected by MAGIC (Teshima et al. 2008). The black dashed line
in the TeV region shows the $\gamma$--ray flux reaching the observer,
considering the absorption through the interaction with the
extragalactic background light estimated with the LowSFR model of
Kneieke et al. (2004) assuming $z=0.31$ for the redshift of
S5 0716+714.  The (negligible) emission from the layer at
small angles is shown by the dotted line.}
\label{3c66b}
\end{figure}

The spectral energy distribution (SED) of the core of 3C66B is shown
in Fig.\ref{3c66b}. We use the equivalent isotropic luminosities
$L_{\nu }$, calculated from the corresponding fluxes $F_{\nu}$ with
$\nu _{\rm r} L_{\nu _{\rm r}}=4\pi d_L^2 \nu _{\rm o}F_{\nu _{\rm
o}}$ (where $d_L$ is the luminosity distance) and the frequency $\nu
_{\rm r}$ is in the rest frame of the source, related to the observed
one by $\nu _{\rm r}= (1+z) \nu _{\rm o}$.  The TeV spectrum (red
open squares) is taken from Aliu et al. (2008) and it has been
corrected (blue open squares) for the (small) attenuation using the
LowSFR model of Kneiske et al. (2004) which predicts a low level of
the extragalactic background light close to what is presently inferred
from observations (e.g. Franceschini et al. 2008) and indirect
arguments (Aharonian et al. 2006b, Mazin \& Raue 2007, Albert et
al. 2008b).

In the X--rays this radiogalaxy shows a rather peculiar behaviour,
with the spectrum apparently changing from soft to very hard (Trussoni
et al. 2003, Balmaverde, Capetti \& Grandi 2006, Evans et
al. 2006). In particular, comparing {\it Chandra} and {\it ROSAT}
observations, Trussoni et al. (2003) concluded that there are hints
that the X--ray flux from the core of the source decreased by a factor
$\sim 4$ in 10 years with an important softening of the spectrum.
This behaviour, unusual for a radiogalaxy, is reminiscent of that of
blazars and could support the idea that the emission comes from the
misaligned jet. However, note that, due to the complexity of the
observed X-ray spectrum, requiring several spectral components, a
strong conclusion is not possible. Optical-UV data for the core have
been obtained with the Hubble Space Telescope (Chiaberge, Celotti \&
Capetti 1999).

There is large consensus that the radio, the optical-UV and, possibly,
the X-ray emission from the core of FRI radiogalaxies is due to
synchrotron emission from the inner jet (Chiaberge et al. 1999, 2002,
Hardcastle \& Worrall 2000, Verdoes Kleijn et al. 2002, Balmaverde et
al. 2006). For 3C66B a direct possibility could therefore be that the
VHE emission is part of the high-energy inverse Compton component from
the same regions of the jet. However, as detailed in TG08, a
single--zone emission model fails in reproducing a SED with two
emission peaks highly separated in frequency: 3C66B, in fact, has the
first peak in the IR and the second peak probably close to TeV
energies. The structured jet model, instead, allows to correctly
reproduce the emission with parameters similar to those commonly
inferred for blazar jets.

We model the SED using the structured jet model of GTC05 and applied
to M87 in TG08. Briefly, we assume that the jet has a inner fast core
(spine), with bulk Lorentz factor $\Gamma_{\rm S}$, surrounded by a
slower layer, with bulk Lorentz factor $\Gamma_{\rm L}$. In both
regions, relativistic electrons emit through synchrotron and inverse
Compton mechanisms.  The existence of a velocity structure impacts on
the observed emission properties of the jet.  Specifically, the
radiative interplay between the layer and the spine amplifies the
inverse Compton emission of both components. Each component sees the
emission of the other amplified because of the relative speed: this
external radiation contributes to the total energy density, enhancing
the emitted inverse Compton radiation.  Depending on the parameters,
this ``external Compton'' (EC) emission can dominate over the internal
synchrotron self--Compton (SSC) component that, especially in TeV
blazars, is depressed because scatterings mainly occur in the
Klein--Nishina (KN) regime. We refer the reader to GTC05 for a
more detailed discussion of the treatment used to calculate the EC
term.

\begin{table*} 
\begin{center}
\begin{tabular}{|l|lllllllllllll|}
\hline
\hline
& $R$    
& $H$  
& $L_{\rm syn}$  
& $n_e^a$
& $B$  
& $\gamma_{\rm min} $  
& $\gamma_{\rm b} $ 
& $\gamma_{\rm max}$  
& $n_1$
& $n_2$
& $\Gamma$ 
& $\theta_{\rm v}$ \\
&cm  &cm &erg s$^{-1}$ & cm$^{-3}$ & G & & & & & & &deg. \\
\hline  
3C66B spine &$5\times 10^{15}$ &$5\times 10^{15}$ &$10^{41}$ & $1.3\times 10^3$ &1.8 &1 &$5\times 10^3$ &$6\times 10^4$   &2   &4   &10  &20 \\ 
3C66B layer &$5\times 10^{15}$ &$2\times 10^{16}$ &$10^{41}$ &1.4 &0.8 &1 &$3\times 10^6$ &$2\times 10^7$ &1.7 &3.2 &3  &20  \\
\hline
M87 spine &$7.5\times 10^{15}$ &$3\times 10^{15}$ &$5.2\times 10^{41}$ & 140.8&1&600 &$5\times 10^3$ &$1\times 10^8$ &2 &3.7 &12 &18 \\ 
M87 layer &$7.5\times 10^{15}$ &$6\times 10^{16}$ &$4.0\times 10^{38}$ & 2.04&0.2 &1&$6\times 10^6$ &$1\times 10^9$ &2 &3.7 &4  &18  \\
\hline
\hline
\end{tabular}                                                         
\caption{Input parameters of the models for the layer and the spine
shown in Fig. \ref{3c66b}. All quantities (except the bulk Lorentz
factors $\Gamma$ and the viewing angle $\theta_{\rm v}$) are measured
in the rest frame of the emitting plasma.  The external radius of the
layer is fixed to the value $R_2=1.2 \times R$. For direct comparison,
we also report the parameters used for M87 (high state, from
TG08). $^a$: density of the relativistic electrons. Note that this is
not a direct input parameter, since it is determined by the assumed
$L_{\rm syn}$.}
\end{center}
\label{tab1}
\end{table*}                                                                  

The model is specified by the following parameters: {\it i)} the spine
is assumed to be a cylinder of radius $R$, height $H_{\rm S}$ (as
measured in the spine frame) and in motion with bulk Lorentz factor
$\Gamma_{\rm S}$; {\it ii)} the layer is modeled as an hollow cylinder
with internal radius $R$, external radius $R_2$, height $H_{\rm L}$
(as measured in the frame of the layer) and bulk Lorentz factor
$\Gamma_{\rm L}$.  Each region contains tangled magnetic field with
intensity $B_{\rm S}$, $B_{\rm L}$ and it is filled by relativistic
electrons assumed to follow a (purely phenomenological) smoothed
broken power--law distribution extending from $\gamma_{\rm min}$ to
$\gamma_{\rm max}$ and with indices $n_1$, $n_2$ below and above the
break at $\gamma_{\rm b}$.  The normalisation of this distribution is
calculated assuming that the system produces an assumed 
(bolometric) synchrotron luminosity $L_{\rm syn}$ (as measured in the
local frame), which is an input parameter of the model.  As in GTC05
and TG08 we assume that $H_{\rm L} > H_{\rm S}$. As said above, the
seed photons for the IC scattering are not only those produced locally
in the spine (layer), but we also consider the photons produced in the
layer (spine).

In Fig. \ref{3c66b} (lower part) we report the model reproducing the
data assuming the emission from the spine (red), accounting for the
data from the radio to the optical--UV bands, and the layer (blue),
assumed to emit the VHE radiation. In view of the possible
contribution of 3C66A to the observed MAGIC spectrum (see Sect.3), we
assume that the emission from the layer of 3C66B lies (slightly) below
the point at the lowest energy. Parameters for the spine and the layer
are reported in Table 1. To produce TeV photons, electrons of the 
layer must have a high value of $\gamma_{\rm b}$. Consequently, for
typical values of the magnetic field intensity, the corresponding
synchrotron emission peaks in the X-ray band, determining a hard X-ray
spectrum. The SSC emission from the layer is negligible, since most of
the scatterings happen deeply in the KN regime.  Therefore the
high--energy emission of the layer, accounting for the measured MAGIC
spectrum, is dominated by the EC component.  On the contrary, the EC
emission from the spine does not contribute significantly to the
emission (at the viewing angles considered here: see the discussion in
Section 2.3) and the high--energy bump is dominated by the SSC
process. 
Since the spine has to reproduce the optical-UV peak the
corresponding value of $\gamma_{\rm b}$ is lower than what assumed for
the layer.

With the adopted parameters and assuming 1 proton per relativistic
electrons (see e.g. Celotti \& Ghisellini 2008) the estimated energy
flux is $4.5\times10^{44}$ erg s$^{-1}$ for the spine (dominated by
protons), and $2.5\times10^{41}$ erg s$^{-1}$ for the layer
(magnetically dominated).  A value around few $10^{44}$ erg s$^{-1}$
is within the range of energy fluxes estimated for FRIs (e.g. Bicknell
\& Begelman 1996, Laing \& Bridle 2002).  The value of the energy flux
for the spine, largely dominated by protons, depends on their number
density, in turn related to $\gamma_{\rm min}$.  In our fit we used
$\gamma _{\rm min}=50$.  Smaller values of $\gamma _{\rm min}$ do not
affect the shape of the SEDs and in principle could be acceptable,
though the derived energy flux would larger.  In the the spine the
ratio of Poynting flux to particle energy flux is dominated (by a
factor $\sim$10) by particles (cold protons).  The layer, instead,
appears to be magnetically dominated, with the Poynting flux about ten
times the particle energy flux.  In the spine, relativistic electrons
and magnetic field are close to equipartition, while in the layer the
magnetic energy density is about two orders of magnitude larger than
the electron energy density.

\subsection{3C66B and M87}

It is interesting to compare the SED and the parameters of the model
obtained for for 3C66B with those of M87 (also reported in Table 1).

The shape of the SED of the core of both radiogalaxies in the radio
and the optical-UV bands appears rather similar, with a peak in the IR
region. Also the luminosities (between $10^{40}$ erg s$^{-1}$ and
$10^{41}$ erg s$^{-1}$) are comparable. In the X-ray band, M87
displays a soft spectrum, close to that of 3C66B as measured by {\it
Chandra}. However, for M87 there are no hints of variability, as in
the case of 3C66B.

A remarkable difference, instead, concerns the importance of the VHE
emission.
While the luminosity of M87 and 3C66B in radio, optical and X--rays
(assumed to be produced by the spine) is comparable,
the VHE luminosity (from the layer) is very different:
even if the flux emitted at 150 GeV is contaminated by 3C66A (see also
below) the output at VHE of 3C66B is more than an order of magnitude larger
than for M87. The difference in the luminosity at TeV energies
translates in the different (comoving) luminosities for the layer
$L_{\rm syn}$ used in our model: $10^{41}$ erg s$^{-1}$ for 3C66B and
$10^{38}$ erg s$^{-1}$ for M87. In 3C66B the layer requires the same
power output of the spine. 
All the other parameters are instead very similar.
The value of $\gamma _{\rm max}$, different between
the two sources, is physically unimportant, as long as 
$\gamma _{\rm max}\gg \gamma_{\rm b}$. 
Of course, these values refer to the specific
(probably high) level of the emission of 3C66B at the epoch of the
observations of MAGIC. Lower luminosities in the TeV band are
possible. Indeed, in our scenario important variations of the emission
from the layer are required to reproduce the variations observed in
the X-ray band.

\subsection{Pairing 3C66B and S5 0716+714}

One can wonder what would be the SED calculated by our model for an
observer located at a small viewing angle. Since we ought to obtain a
SED similar to the SED of known blazars, this should be considered as
a consistency check for our model and for the derived parameters.  In
Fig. \ref{3c66b} we report the predicted SED for a viewing angle of
$2$ degrees and compare it to the multi--frequency observations of the
BL Lac S5 0716+714 at the redshift $z=0.31$ (Witzel et al. 1988; but
possibly larger, Sbarufatti, Treves \& Falomo 2005).  This blazar has
been recently detected in the TeV band by MAGIC (Teshima et al. 2008).
Although we do not pretend that the model exactly reproduces the data,
one can see that the two are rather similar. As discussed in TG08,
the condition that the SED at small angles resembles those of known
blazars dictates the choice that the VHE radiation comes from the
layer.

At small angles, the emission is entirely dominated by the spine,
which is characterised by a larger Lorentz factor.  The direct
emission from the layer (dotted line) provides a negligible
contribution to the total observed emission.  An important point to
note is the rather hard spectrum predicted in the GeV band. The
emission in this spectral band is mainly due to the EC component of
the spine Comptonizing the seed photons provided by the layer.  For
comparison, the SSC component of the spine is reported by the dashed
orange line. As already noted, the EC component from the spine is not
prominent at larger angles (note the different shape of the high
energy component of the spine for $\theta_{\rm v}=20$ deg). The reason
is that, contrary to the case of the layer, the EC beaming cone of the
spine is narrower than the corresponding synchrotron and SSC cones, an
effect originally pointed out by Dermer (1995) in discussing the
external inverse Compton emission in blazars. This effect is due to
the fact that the seed photons produced in the layer are seen, in the
spine frame, as coming from the forward direction.  The pattern of the
scattered radiation, even in the spine frame, is therefore {\it
anisotropic}: more power is produced in directions close to the jet
axis.  When transforming in the observer frame, this translates into a
radiation pattern that is more enhanced along the velocity (and jet
axis) direction than the one usually derived when dealing with
radiation that is isotropic in the comoving frame (see Fig.2 of TG08
that illustrates this point).

The presence of the layer is thus important even when the source is
observed at small viewing angles and the entire radiation is provided
by the spine. It enhances the observed radiation produced by the spine
at high energies (GeV band), just where the suppression of the SSC
flux due to KN effects starts to be important. The contribution of the
seed photons of the layer is important when the layer luminosity is
large, as in the case of 3C66B.  This is the reason why, in the case
of M87 paired with BL Lac, it was not noticeable: the layer
luminosity, in that case, was much smaller.

\section{The VHE spectra of 3C66A/B}

We cannot exclude that the nearby 3C66A contributes to the
total emission measured by MAGIC, also in view of the recent detection
by VERITAS (Swordy et al. 2008). However, due to the (uncertain) high
redhift of this blazar, a sizeable contribution to the measured spectrum is
expected only at the lowest energies, $\sim 100$ GeV.

\begin{figure}
\vskip -0.5 true cm
\centerline{ 
\psfig{file=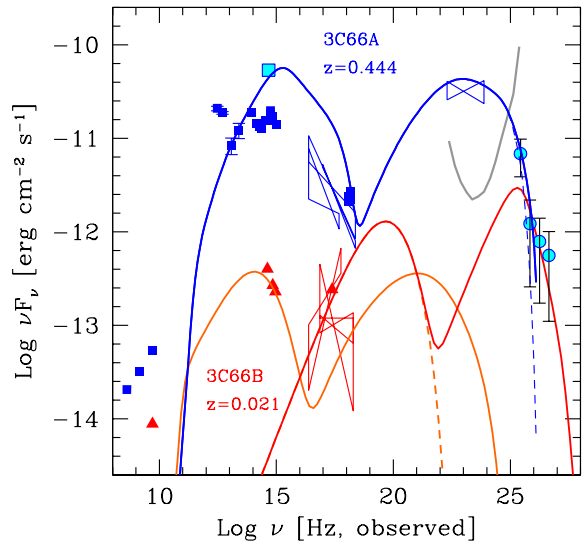,height=8.5cm,width=8.5cm} }
\vskip -0.5 true cm
\caption{SED of 3C66B as in Fig. \ref{3c66b} and of 3C66A. Data for
3C66A (blue squares and bow-ties) are from NED (radio, optical), Impey
\& Neugebauer (1988) (IRAS), Donato et al. (2001) (ROSAT), Donato et
al. (2005) ({\it Beppo}SAX), Foschini et al. (2006) (XMM--Newton), Hartman et
al. (1999) (EGRET). The square in cyan reports the average optical
($R$--band) flux during the observations of MAGIC. The SED of 3C66A has
been reproduced with the model for the blazar emission described in
Celotti \& Ghisellini (2008). The solid line is the spectrum emitted
at the source, while the dashed line reports the $\gamma$--ray
emission reaching the Earth after being absorbed by the interaction
with the extragalactic background light (we assumed the LowSFR model
of Kneiske et al. 2004 and a redshift of $z=0.444$).  The measured VHE
spectrum of MAGIC (cyan circles) is reproduced as the sum of the
emission from both sources, 3C66A being dominant below 200 GeV, and
3C66B accounting for the emission above 200 GeV. The grey line reports
the {\it Fermi} sensitivity (for 1 year and 5$\sigma$ significance). 
}
\label{3c66ab}
\end{figure}

The SEDs of 3C66A and B are compared in Fig. \ref{3c66ab}. For 3C66A we
used historical data (see figure caption for references), together
with the $R$-band flux averaged over the period of observation of
MAGIC (Aliu et al. 2008). Unfortunately there are no observations in
the GeV band of 3C66A around the epoch of the MAGIC observations. 
Therefore, in our model we assume the average spectrum (bow-tie)
reported in the 3rd EGRET catalogue (Hartman et al. 1999). 

We quantify the possible contribution of 3C66A to the spectrum
measured by MAGIC modeling the SED, with the model for blazars
described in Celotti \& Ghisellini (2008)\footnote{We use the
following parameters: $R=1.5\times 10^{16}$ cm, $\Delta R= R$,
$L'=3.8\times 10^{42}$ erg s$^{-1}$, $B= 1.4$ G, $\gamma_{\rm min}=
1$, $\gamma_{\rm b} = 4.5\times 10^3$, $\gamma_{\rm max} = 10^5$,
$n_2=3.6$, $\Gamma=18$, $\theta_v = 3$ deg. See Celotti and Ghisellini
(2008) for a full description of these parameters.}. To compare the
data with the model, the curve of the intrinsic emission (solid line)
has been absorbed with the LowSFR model of Kneiske et al. (2004)
(dashed line). Due to the severe absorption, the flux from 3C66A above
100 GeV is strongly depressed and only 3C66B contributes to the
observed spectrum. However, 3C66A could entirely account for the flux
measured at the lowest energies. In this case, the contribution of
3C66A alleviates the requirement on the luminosity of the layer of
3C66B by a factor of 3 (as already noted the line for our model of
3C66B stays below the first bin of the MAGIC spectrum).

Of course, since no simultaneous data (especially at high energy) are
available for 3C66A, our model of the SED is somewhat arbitrary. In
particular, despite the strong absorption, a detectable emission in
the TeV band is expected in states of high activity of the source. In
fact, the recent detection by VERITAS, with a flux above $10^{-11}$
erg cm$^{-2}$ s$^{-1}$ (Swordy et al. 2008), has been obtained during
a period of rather intense $\gamma $--ray activity, detected by {\it
Fermi} (Tosti et al. 2008).

\section{Discussion}

The possible detection of VHE photons from the FRI radiogalaxy 3C66B
(at a distance of 85.5 Mpc) suggests that M87 is not an isolated case
but, instead, radiogalaxies could represent a new class of
extragalactic TeV sources. The fact that radiogalaxies can be
relatively luminous $\gamma $--ray sources was proposed by GTC05 based
on the idea of the existence of structured jets. According to this
scenario, in sources with misaligned jets, the emission from the fast
regions of the jet is deboosted while the slower layer can still
provide an important contribution because of the broader emission
cone.

The best VHE candidates are the radiogalaxies in which the jet axis
lies at a relatively small angle with respect to the line of
sight. M87, with an angle between 30 and 20 degrees is probably one of
the best cases. For 3C66B we assume in our model a similar angle, 20
degrees. For larger angles the deboosting becomes more important and
thus the power requirement grows substantially. The consequent
increase of the intrinsic luminosity of the spine in the optical-IR
band results in a large optical depth for the pair production process
for VHE photons within the layer.
Radio VLBI observations (Giovannini et al. 2001) show a complex
structure for 3C66B, with the possible presence of a counter-jet at
small distances, disappearing at larger distances. Giovannini et
al. (2001) interpreted this behavior as the evidence that the jet
accelerates, from $\beta=0.6$ at 1.5 mas (linear deprojected distance
of about 1 parsec) to $\beta=0.99$ at 4.5 mas from the core and a jet
inclination of about 45 degrees. However, this interpretation is not
unique, the same effect being produced by a jet with constant speed,
$\beta=0.83$, varying the angle (of both the jet and the counter-jet)
from $\theta =60$ deg to $\theta =30$ deg. Moreover, this conclusion
is based on the implicit assumption that the jets are exactly
symmetric. Local variations of the emissivity in one of the two jets
could easily mimic such a phenomenology. Given these uncertainties we
consider that $\theta =20$ degrees for the inner jet (at distances of 0.1
pc) is possible.

As extensively discussed in TG08, the model parameters are not
completely constrained by the data. A whole family of solutions is
allowed.  However, the requirement that the parameters of the spine,
and its SED at small angles, are similar to those of blazars allows us
to partly constrain the parameters.

In our model we implicitly assume that the region emitting
$\gamma$-rays from 3C66B is, as for M87, the inner jet. However, since
no variations have been detected, the alternative models invoking
emission from more distant regions along the jet (e.g. Stawarz, Sikora
\& Ostrowski 2003) cannot be excluded. Further observations are needed
to clarify this point.

An important point to note is the important contribution of the EC
component to the spine emission at small angles. The contribution is
particular relevant above the GeV band, determining a hard spectrum up
to the VHE band. In the previous model for M87 this component was not
so prominent due to the lower intrinsic luminosity required for the
layer. In BL Lacs this component could play an important role in
determining the high-energy $\gamma$-ray spectrum. In particular,
since SSC dominates below and the EC above the GeV band, variability
of the layer emission could determine different variability of the
emission in the two bands. Simultaneous observations with {\it Fermi}
and Cherenkov telescopes could test this possibility.

Another important point to mention is that, due to its limited angular
resolution, {\it Fermi} cannot distinguish between the two sources,
3C66A and B. However, the MeV-GeV band should be largely dominated by
3C66A, with a rather small contribution from the radiogalaxy only at
the largest energies covered by {\it Fermi}. Above 100 GeV, however,
the situation reverses, and thus the emission can be dominated by
3C66B when 3C66A is at low level. In these occasions, Cherenkov
telescopes, characterized by a good angular resolution, could possibly
simultaneously detect both sources.

\section*{Acknowledgements}
We thank L. Maraschi, D. Mazin, M. Persic and R. Wagner for useful
comments and M. Chiaberge and E. Trussoni for discussions. We are
grateful to the referee, G. Bicknell, for the constructive and useful
report. This research has made use of the NASA/IPAC Extragalactic
Database (NED) which is operated by the Jet Propulsion Laboratory,
California Institute of Technology, under contract with the National
Aeronautics and Space Administration. This work was partly financially
supported by a 2007 COFIN-MiUR grant.

\end{document}